\documentclass[aip, cha, 11pt]{revtex4}

\pdfoutput=1
\usepackage{graphicx}
\usepackage{amsmath}
\usepackage{amssymb}
\usepackage{bm}

\newcommand{\pa}{\partial}

\newcommand{\mean}[1]{\langle{#1}\rangle}

\newcommand{\bra}[1]{\langle{#1}|}
\newcommand{\ket}[1]{|{#1}\rangle}

\newcommand{\Tr}{{\rm Tr}\hspace{0.07cm}}

\newcommand{\abs}[1]{{|#1|}}

\begin{document}

\title{
A definition of the asymptotic phase for quantum nonlinear oscillators from the Koopman operator viewpoint
}

\author{Yuzuru Kato}
\thanks{Corresponding author. E-mail: katoyuzu@fun.ac.jp}
\affiliation{Department of Complex and Intelligent Systems, Future University Hakodate, Hokkaido 041-8655, Japan }

\author{Hiroya Nakao}
\thanks{E-mail: nakao@sc.e.titech.ac.jp}
\affiliation{Department of Systems and Control Engineering, Tokyo Institute of Technology, Tokyo 152-8552, Japan}

\date{\today}

\begin{abstract}
We propose a definition of the asymptotic phase for quantum nonlinear oscillators from the viewpoint of the Koopman operator theory. The asymptotic phase is a fundamental quantity for the analysis of classical limit-cycle oscillators, but it has not been defined explicitly for quantum nonlinear oscillators. In this study, we define the asymptotic phase for quantum oscillatory systems by using the eigenoperator of the backward Liouville operator associated with the fundamental oscillation frequency. By using the quantum van der Pol oscillator with Kerr effect as an example, we illustrate that the proposed asymptotic phase appropriately yields isochronous phase values in both semiclassical and strong quantum regimes.
\end{abstract}

\maketitle


{\bf 
Spontaneous rhythmic oscillations and synchronization are observed in a wide variety of classical rhythmic systems, and recent progress in nanotechnology is facilitating the analysis of quantum rhythmic systems.
The asymptotic phase plays a fundamental
role in the analysis of classical limit-cycle oscillators,
but a fully quantum-mechanical definition for quantum limit-cycle oscillators has been lacking.
In this study, we propose a definition of the asymptotic phase for quantum nonlinear oscillators, which naturally extends the definition of the asymptotic phase for classical stochastic oscillatory systems~\cite{thomas2014asymptotic} from the Koopman-operator viewpoint~\cite{kato2021asymptotic}
and provides us with appropriate phase values for characterizing quantum synchronization.
}


\section{Introduction}

Synchronization of spontaneous rhythmic oscillations are widely observed in nature~\cite{winfree2001geometry, kuramoto1984chemical, 
	pikovsky2001synchronization, nakao2016phase, ermentrout2010mathematical, strogatz1994nonlinear}.
It has been extensively studied in a variety of classical systems in physics, chemistry, and biology.
Recently, much progress has been made in the experimental realization 
of synchronization in micro- and nanoscale systems, such as 
nanoelectromechanical oscillators \cite{matheny2019exotic}, micro lasers \cite{kreinberg2019mutual}, spin torque oscillators \cite{singh2019mutual},
and optomechanical oscillators \cite{colombano2019synchronization}.
Stimulated by the experimental developments,
quantum synchronization has attracted much attention recently
\cite{
	lee2013quantum,
	walter2014quantum, 
	sonar2018squeezing,
	kato2019semiclassical,
	mok2020synchronization, 
	lee2014entanglement,
	witthaut2017classical,
	roulet2018quantum,
	lorch2016genuine,
	lorch2017quantum,
	nigg2018observing,
	weiss2016noise,
	es2020synchronization,
	kato2021enhancement, 
	kato2021instantaneous,
	li2021quantum, 
	laskar2020observation, 
	koppenhofer2020quantum,
	kato2020semiclassical,
	hush2015spin,
	weiss2017quantum,
	mari2013measures,
	xu2014synchronization,
	roulet2018synchronizing,
	chia2020relaxation, 
	arosh2021quantum, 
	jaseem2020generalized, 
	jaseem2020quantum, 
    cabot2021metastable}
and theoretical investigations of quantum signatures in synchronization, such as quantum fluctuations~\cite{lee2013quantum, walter2014quantum, sonar2018squeezing, kato2019semiclassical,  kato2020semiclassical},
quantum entanglement~\cite{lee2014entanglement, witthaut2017classical, roulet2018quantum}, discrete nature of the energy spectrum~\cite{lorch2016genuine, lorch2017quantum, nigg2018observing},  
and effects of quantum measurement~\cite{weiss2016noise, es2020synchronization, kato2021enhancement, kato2021instantaneous, li2021quantum}, have been carried out.
Experimental realizations of quantum phase synchronization in spin-$1$ atoms \cite{laskar2020observation}
and on the IBM Q system \cite{koppenhofer2020quantum} have also been reported recently.

In classical deterministic systems, spontaneous rhythmic oscillations are typically modeled as stable limit cycles of nonlinear dynamical systems.
The \textit{asymptotic phase}~\cite{winfree2001geometry, kuramoto1984chemical, pikovsky2001synchronization, nakao2016phase, ermentrout2010mathematical}, defined by the oscillator's vector field and increases with a constant frequency in the basin of the limit cycle, plays a central role
in analyzing synchronization properties of limit-cycle oscillators.
It is the basis for \textit{phase reduction}~\cite{winfree2001geometry, kuramoto1984chemical, pikovsky2001synchronization, nakao2016phase, 	ermentrout2010mathematical, strogatz1994nonlinear}, 
which gives low-dimensional phase equations approximately describing the oscillators under weak perturbations.
Recently, it has been clarified that the asymptotic phase, which was originally introduced from a geometrical viewpoint~\cite{winfree2001geometry}, has a natural relationship with the Koopman eigenfunction associated with the fundamental frequency of the oscillator~
\cite{mauroy2013isostables, mauroy2020koopman, shirasaka2017phase, kuramoto2019concept, kato2021asymptotic}.

For classical stochastic oscillatory systems, Thomas and Lindner~\cite{thomas2014asymptotic} proposed a definition of the asymptotic phase in terms of the slowest decaying eigenfunction of the backward Fokker-Planck (Kolmogorov) operator describing the mean first passage time, which appropriately yields isochronous phase values that increase with a constant frequency on average even for strongly stochastic oscillations, in a similar way to the ordinary asymptotic phase for deterministic oscillators.
We recently pointed out that their definition can be considered a natural extension of the deterministic definition from the Koopman operator viewpoint~\cite{kato2021asymptotic} (see \cite{junge2004uncertainty, vcrnjaric2020koopman, wanner2020robust} for the details of the stochastic Koopman operator).

The classical definitions of the asymptotic phase are applicable to quantum nonlinear oscillators
in the semiclassical regime, where the system is described by a stochastic differential equation for the phase-space state along a deterministic classical trajectory under the effect of small quantum noise~\cite{kato2019semiclassical,kato2020semiclassical, kato2021asymptotic}.
However, in the stronger quantum regime, we cannot rely on the semiclassical approximation and how to define the asymptotic phase in a fully quantum-mechanical manner is an open question.
In this study, we propose a definition of the asymptotic phase for quantum nonlinear oscillatory systems by using the eigenoperator of the adjoint Liouville operator, which is a counterpart of the backward Fokker-Planck operator in classical stochastic systems. We illustrate the validity of our definition by using a quantum van der Pol oscillator with quantum Kerr effect in both semiclassical and strong quantum regimes. 

\section{Asymptotic phase for classical oscillatory systems}

In this section, we briefly review the definitions of the asymptotic phase for deterministic~\cite{winfree2001geometry, kuramoto1984chemical, pikovsky2001synchronization, nakao2016phase, ermentrout2010mathematical, mauroy2020koopman, shirasaka2017phase, kuramoto2019concept} and stochastic~\cite{thomas2014asymptotic,kato2021asymptotic} classical oscillatory systems.

\subsection{Deterministic oscillatory systems}

Consider a deterministic dynamical system 
\begin{align}
	\dot {\bm X}(t) = \bm{A}(\bm{X}(t)),
	\label{eq1}
\end{align}
where $\bm{X}(t) \in \mathbb R^{N}$ is the system state at time $t$, ${\bm A}({\bm X}) \in {\mathbb R}^{N}$ is a vector field representing the system dynamics, and $(\dot{})$ represents time derivative.
We assume that this system has an exponentially stable limit-cycle solution ${\bm{X}}_{0}(t)$ with a natural period $T$ and frequency $\Omega_c = 2\pi / T$, satisfying ${\bm X}_0(t+T) = {\bm X}_0(t)$, and denote its basin of attraction as $B \subseteq {\mathbb R}^N$.
The asymptotic phase  
$\Phi_c({\bm{X}}) : {B} \to [0, 2\pi)$
is defined such that 
${\bm A}({\bm{X}})  \cdot  \nabla \Phi_c({\bm{X}}) = \Omega_c$
is satisfied for 
$\forall {\bm X} \in B$,
where $\nabla = \partial / \partial {\bm X}$ represents the gradient with respect to ${\bm X}$~\cite{winfree2001geometry, kuramoto1984chemical, pikovsky2001synchronization, nakao2016phase, ermentrout2010mathematical}.
The asymptotic phase $\phi(t) = \Phi_c({\bm X}(t))$ of the system state ${\bm X}(t)$ then obeys
\begin{align}
	\dot{\phi}(t) = \dot{\Phi}_c({\bm X}(t)) = {\bm A}({\bm X}(t)) \cdot \nabla \Phi_c({\bm X}(t)) = \Omega_c,
	\label{asymphase2}
\end{align}
i.e., $\phi$ always increases with a constant frequency $\Omega_c$ as ${\bm X}$ evolves in $B$. 
Thus, the asymptotic phase $\Phi_c$ gives a nonlinear transformation of the system state ${\bm X}$ to a phase value $\phi$ such that the dynamics of $\phi$ takes a simple linear form, $\phi(t) = \Omega_c t + const.$
The simplicity of the phase equation~(\ref{asymphase2}) has facilitated detailed studies of synchronization and collective dynamics in coupled-oscillator systems~\cite{winfree2001geometry, kuramoto1984chemical, pikovsky2001synchronization, nakao2016phase, ermentrout2010mathematical}.
The level sets of $\Phi_c({\bm X})$ are called isochrons.

The linear operator
$	{A} = {\bm A}({\bm X}) \cdot \nabla$
in the definition of the asymptotic phase $\Phi_c({\bm X})$ is the infinitesimal generator of the {\it Koopman operator} describing the evolution of observables for the system described by Eq.~(\ref{eq1}) (see Refs.~\cite{mauroy2013isostables, mauroy2020koopman, shirasaka2017phase, kuramoto2019concept} for details).
The complex exponential 
$	\Psi_c({\bm X}) = e^{i \Phi_c({\bm X})} $
of $\Phi_c({\bm X})$ is an eigenfunction of ${A}$ with the eigenvalue $i \Omega_c$, namely, it satisfies the eigenvalue equation
$	{A} \Psi_c({\bm X}) = i \Omega_c \Psi_c({\bm X})$.
Therefore, the asymptotic phase $\Phi_c({\bm X})$ has a natural operator-theoretic interpretation as the argument of the Koopman eigenfunction $\Psi_c({\bm X})$ associated with the eigenvalue $i \Omega_c$, characterized by the fundamental frequency $\Omega_c$ of the oscillator~\cite{mauroy2013isostables, mauroy2020koopman, shirasaka2017phase, kuramoto2019concept}.
We can thus define the asymptotic phase by using the Koopman eigenfunction $\Psi_c({\bm X})$ of $A$ as
\begin{align}
	\label{eq:koopmaniw}
	\Phi_c({\bm X}) = \arg \Psi_c({\bm X}).
\end{align}

\subsection{Stochastic oscillatory systems}

For stochastic oscillatory systems, we cannot use the deterministic limit-cycle solution in defining the asymptotic phase unless the noise is sufficiently weak.
Thomas and Lindner~\cite{thomas2014asymptotic} defined the asymptotic phase for classical stochastic oscillators without relying on the deterministic limit cycle by using the eigenfunction with the slowest decay rate of the backward Fokker-Planck operator.

Consider a stochastic oscillator described by an Ito stochastic differential equation (SDE)
\begin{align}
	d{\bm X}(t) = {\bm A}({\bm X}(t)) dt + {\bm B}({\bm X}(t)) d{\bm W}(t),
	\label{sde}
\end{align}
where $\bm{X}(t) \in \mathbb R^{N}$ is the system state at time $t$, ${\bm A}({\bm X}) \in {\mathbb R}^N$ is a drift vector representing the deterministic dynamics, ${\bm B}({\bm X}) \in {\mathbb R}^{N \times N}$ is a matrix characterizing the effect of the noise, and ${\bm W}(t) \in {\mathbb R}^N$ is a $N$-dimensional Wiener process.
This system is assumed to be {\it oscillatory} in the sense explained below.
The transition probability density
$p(\bm {X}, t| \bm {Y}, s)$ ($t \geq s$)
of Eq.~(\ref{sde}) obeys the forward and backward Fokker-Planck equations~\cite{gardiner2009stochastic},
\begin{align}
	\frac{\pa}{\pa t} p(\bm {X}, t| \bm {Y}, s) = {L}_{\bm X}  p(\bm {X}, t| \bm {Y}, s),
	\label{forward}
\end{align}
and
\begin{align}
	\frac{\pa}{\pa s} p(\bm {X}, t| \bm {Y}, s) &= -{L}^*_{\bm Y} p(\bm {X}, t| \bm {Y}, s),
\end{align}
respectively, where the (forward) Fokker-Planck operator is given by
\begin{align}
	\label{eq:fpe2}
	{L}_{\bm X} = - \frac{\pa}{\pa {\bm X}} {\bm A}({\bm X}) + \frac{1}{2} \frac{\pa^2}{\pa {\bm X}^2} {\bm D}({\bm X})
\end{align}
and the backward Fokker-Planck operator is given by (in terms of ${\bm X}$)
\begin{align}
	{L}_{\bm X}^* = {\bm A}({\bm X}) \frac{\pa}{\pa {\bm X}} + \frac{1}{2} {\bm D}({\bm X}) \frac{\pa^2}{\pa {\bm X}^2}.
	\label{eq:Lxadj}
\end{align}
Here, ${\bm D}({\bm X}) = {\bm B}({\bm X}) {\bm B}({\bm X})^{\sf T} \in {\mathbb R}^{N \times N}$ is a matrix of diffusion coefficients (${\sf T}$ indicates matrix transposition).
The forward and backward operators $L_{\bm X}$ and $L_{\bm X}^*$ are mutually adjoint, i.e., 
$	\mean{ {L}_{\bm{X}} G(\bm{X}), H(\bm{X})}_{\bm{X}}
	=\mean {G(\bm{X}), {L}^*_{\bm X} H(\bm{X})}_{\bm{X}}$,
where the inner product is defined as
$	\mean{G(\bm{X}), H(\bm{X})}_{\bm{X}} = \int \overline{ G(\bm{X}) } H(\bm{X}) d \bm{X}$
for two functions $G(\bm{X}), H(\bm{X}) : {\mathbb R}^N \to {\mathbb C}$ (the overline indicates complex conjugate and the integration is taken over the whole range of ${\bm X}$).

As explained in Appendix A, the backward Fokker-Planck operator $L_{\bm X}^*$ in Eq.~(\ref{eq:Lxadj}) is the infinitesimal generator of the Koopman operator for Eq.~(\ref{sde}).
We denote the probability density function of ${\bm X}$ at time $t$ as $p_t({\bm X}) \in {\mathbb R}$, which also obeys the Fokker-Planck equation $\partial p_t({\bm X}) / \partial t = L_{\bm X} p_t({\bm X})$, and an observable of the system at time $t$ as $g_t : {\mathbb R}^N \to {\mathbb C}$, which maps the system state ${\bm X}$ to a complex value.
The evolution of the expectation $\langle g \rangle_t = \int p_t({\bm X}) g_t({\bm X}) d{\bm X} =
 \langle p_t({\bm X}), g_t({\bm X}) \rangle_{ \bm{X}}$of the observable $g$ at $t = t_0$ can be expressed as 
\begin{align}
\left. \frac{d}{dt} \langle g \rangle_t \right|_{t=t_0}
&=
\left \langle \left. \frac{\partial}{\partial t} p_t({\bm X}) \right|_{t = t_0},\  g_{t_0}({\bm X}) \right \rangle_{\bm X}
=
\langle L_{\bm X} p_{t_0}({\bm X}),\ g_{t_0}({\bm X}) \rangle_{\bm X}
\cr
&=
\langle p_{t_0}({\bm X}),\ L_{\bm X}^* g_{t_0}({\bm X}) \rangle_{\bm X}
=
\left \langle  \left. p_{t_0}({\bm X}),\ \frac{\partial}{\partial t} { g_t({\bm X}) }\right|_{t=t_0}  \right \rangle_{\bm X},
\label{clevobs}
\end{align}
where $g_t({\bm X})$ remains constant and $p_t({\bm X})$ evolves in the second expression, while $p_t({\bm X})$ remains constant and $g_t({\bm X})$ evolves in the last expression.

The linear differential operators ${L}_{\bm X}$ and ${L}_{\bm X}^*$ have a biorthogonal eigensystem $\{\lambda_{k}, P_{k}, Q_{k}\}_{k=0, 1, 2, ...}$ of the eigenvalue $\lambda_{k}$ and eigenfunctions $P_k({\bm X})$ and $Q_k({\bm X})$ satisfying
\begin{align}
	{L}_{\bm X} P_{k}(\bm{X}) = \lambda_k P_{k}(\bm{X}),~~
	{L}_{\bm X}^* Q^{}_{k}(\bm{X}) = \overline{ \lambda_k } Q_{k}(\bm{X}),~~
\mean{ P_{k}(\bm{X}), Q_{l}(\bm{X})}_{\bm{X}} 
	= \delta_{kl},
\end{align}
where $k, l = 0, 1, 2, \ldots$ and $\delta_{kl}$ represents the Kronecker delta~\cite{gardiner2009stochastic}.
We assume that, among the eigenvalues, one eigenvalue $\lambda_0$ is zero,  which is associated with the stationary state $p^S({\bm X})$ of the system satisfying $L_{\bm X} p^S({\bm X}) = 0$, and all other eigenvalues have negative real parts. 
It is assumed that the eigenvalues with the largest non-negative real part (or the smallest absolute real part, i.e., the slowest decay rate) are given by a complex-conjugate pair.
We denote these eigenvalues as 
\begin{align}
	\lambda_1 = \mu_s - i \Omega_s,
	\quad
	\overline{ \lambda_1 } = \mu_s + i \Omega_s,
\end{align}
where $\abs{\mu_s}~(\mu_s < 0)$ is the decay rate and $\Omega_s = \mbox{Im}\ \overline{ \lambda_1 }$ represents the fundamental oscillation frequency of the system. The oscillatory property of the system is embodied in this assumption.
We also assume that this pair of principal eigenvalues are well separated from other branches of eigenvalues and this oscillatory mode is dominant in the system.

Thomas and Lindner~\cite{thomas2014asymptotic} proposed a definition of the {\it stochastic asymptotic phase} of the system described by Eq.~(\ref{sde}) by using the argument of the complex conjugate of the eigenfunction 
${ Q_1({\bm X}) }$ of $L^*_{\bm X}$ associated with $\overline{\lambda_1}$ as (in the notation used here)
\begin{align}
	\Phi_s(\bm{X}) = \arg { Q_1 (\bm{X}) }.
	\label{eq:TLphase}
\end{align}
This definition is natural from the Koopman operator viewpoint~\cite{kato2021asymptotic}, as $L_{\bm X}^*$ is the infinitesimal generator of the Koopman operator for Eq.~(\ref{sde}) that formally goes back to the linear operator $A = {\bm A}({\bm X}) \cdot \nabla$, i.e., to the infinitesimal generator of the Koopman operator for the deterministic system Eq.~(\ref{eq1}) in the noiseless limit ${\bm D}({\bm X}) \to 0$.
The exponential average of $\Phi_s$ satisfies~\cite{kato2021asymptotic}
\begin{align}
\frac{d}{dt} \mbox{arg}\ \mathbb{E}^{{\bm X}_0}[ e^{ i \Phi_s({\bm X}(t)) } ] 
= \frac{d}{dt} \mbox{arg}\ \mathbb{E}^{{\bm X}_0}[ Q_1({\bm X}(t)) ]
= \mbox{Im} \overline{\lambda_1} = \Omega_s,
\label{expaveg}
\end{align}
where $\mathbb{E}^{{\bm X}_0}$ represents the average over the stochastic trajectories of Eq.~(\ref{sde}) starting from an initial point ${\bm X}_0 \in {\mathbb R}^N$.
Thus, the asymptotic phase defined by Eq.~(\ref{eq:TLphase}) increases with a constant frequency $\Omega_s$ on average with the stochastic evolution of the system and can be considered a natural generalization of the deterministic asymptotic phase in Eq.~(\ref{eq:koopmaniw}).
It is noted that we may also choose the eigenvalue $\lambda_1$ and eigenfunction $\overline{Q_1 (\bm{X})}$ to define the stochastic asymptotic phase, which reverses its direction of increase.

\section{Asymptotic phase for quantum oscillatory systems}

Our aim in this study is to propose a quantum-mechanical definition of the asymptotic phase that does not rely on classical trajectories.
In Ref.~\cite{kato2019semiclassical}, we developed a semiclassical phase reduction theory for quantum limit-cycle oscillators, but the definition of the asymptotic phase was based on the deterministic limit cycle in the classical limit and could not be applied in stronger quantum regimes. 
In Ref.~\cite{kato2021asymptotic}, we considered the asymptotic phase of quantum limit-cycle oscillators using the definition for the classical stochastic oscillators explained in Sec.~II, but it was valid only in the semiclassical regime.
Here, we consider a quantum master equation describing the evolution of the density operator of the system and define the asymptotic phase by using the eigenoperator of the adjoint Liouville operator.
We use standard notations for open quantum systems without a detailed explanation; see e.g., Refs.~\cite{carmichael2007statistical,gardiner1991quantum,breuer2002theory} for details.

\subsection{Quantum master equation}

We consider quantum oscillatory systems with a single degree of freedom coupled to reservoirs.
The system's quantum state is represented by a Hermitian density operator $\rho$ ($ = \rho^\dag$) and the observable is described by an operator $F$, where $\dag$ represents Hermitian conjugate.
Introducing an inner product $\mean{X,Y}_{tr} =\Tr{ ( X^{\dag}Y )}$ of two operators $X$ and $Y$, the expectation of the observable $F$ is expressed as
\begin{align}
\langle F \rangle = \mbox{Tr}\ ( \rho F ) = \langle \rho, F \rangle_{tr}.
\end{align}

In the Schr\"odinger picture, the quantum state $\rho$ evolves with time while the observable $F$ remains constant.
Assuming that the interactions of the system with the reservoirs are instantaneous and Markovian approximation can be employed, the evolution of the density operator $\rho_t$ at time $t$ obeys the quantum master equation~\cite{carmichael2007statistical,gardiner1991quantum, breuer2002theory}
\begin{align}
	\label{eq:me}
	\dot{\rho_t}
	= {\mathcal L} \rho_t,
\end{align}
where the Liouville operator ${\mathcal L}$ (sometimes called a superoperator because it acts on an operator) is given by
\begin{align}
	{\mathcal L} X =
	-i[H, X] 
	+ \sum_{j=1}^{n} \mathcal{D}[C_{j}] X.
\end{align}
Here, 
$H (=H^{\dag})$ is a system Hamiltonian, 
$C_{j}$ is a coupling operator between the system and 
$j$th reservoir $(j=1,\ldots,n)$,  
$[A,B] = AB-BA$ is the commutator, $\mathcal{D}[C]X = C X C^{\dag} - (X C^{\dag} C + C^{\dag} C X)/2$ 
is the Lindblad form, 
and the reduced Planck's constant is set as $\hbar = 1$.

In the Heisenberg picture, the quantum state $\rho$ remains constant while the observable $F$ evolves with time as
\begin{align}
\dot{F}_t = {\mathcal L}^* F_t,
\end{align}
where the time dependence of $F$ is explicitly shown. Here, ${\mathcal L}^*$ is the adjoint operator of ${\mathcal L}$ satisfying
\begin{align}
 \mean{{\mathcal L} X,Y}_{tr} =\mean{X,{\mathcal L}^* Y}_{tr} 
\end{align}
and can be explicitly calculated as
\begin{align}
	\label{eq:bme}
	{\mathcal L}^* X = i[H, X] + \sum_{j=1}^{n} \mathcal{D}^+[C_{j}]X,
\end{align}
where $\mathcal{D}^+[C]X = C^{\dag} X C - (X C^{\dag} C + C^{\dag} C X)/2$ is the adjoint Lindblad form~\cite{breuer2002theory}.
The evolution of the expectation $\langle F \rangle_t = \langle \rho_t, F_t \rangle_{tr}$ of $F$ with respect to $\rho$ at $t = t_0$ can be expressed as
\begin{align}
\left. \frac{d}{dt} \langle F \rangle_t \right|_{t = t_0}
&= \langle \dot{\rho}_t|_{t=t_0},\ F_{t_0} \rangle_{tr}
= \langle {\mathcal L} {\rho}_{t_0}, F_{t_0} \rangle_{tr} 
\cr
&= \langle {\rho}_{t_0}, {\mathcal L}^* F_{t_0} \rangle_{tr}
= \langle \rho_{t_0}, \dot{F}_t|_{t=t_0} \rangle_{tr},
\label{qmevobs}
\end{align}
where $F$ remains constant and $\rho$ evolves in the second expression (Schr\"odinger picture), while $\rho$ remains constant and $F$ evolves in the last expression (Heisenberg picture).
Equation~(\ref{qmevobs}) corresponds to Eq.~(\ref{clevobs}) for the expectation of the observable for classical stochastic systems.
Thus, the adjoint operator ${\mathcal L}^*$ is a counterpart of the backward Fokker-Planck operator $L_{\bm X}^*$ in Sec.~II, namely, ${\mathcal L}^*$ corresponds to the infinitesimal generator of the Koopman operator.

We assume that the operators ${\mathcal L}$ and ${\mathcal L}^*$ have a biorthogonal eigensystem $\{ \Lambda_{k}, U_{k}, V_{k} \}_{k=0, 1, 2, ...}$ consisting of the eigenvalue $\Lambda_k$ and right and left eigenoperators $U_k$ and $V_k$, satisfying
\begin{align}
	{\mathcal L} U_{k} =  \Lambda_{k} U_{k},
	\quad
	{\mathcal L}^* V_{k} =  \overline{\Lambda_{k}} V_{k},
	\quad
    \mean{U_{k}, V_{l}}_{tr} = \delta_{kl}, 
	\label{eq:eigentriplet1}
\end{align}
for $k, l=0, 1, 2, \ldots$~\cite{li2014perturbative}.
Among the eigenvalues, one eigenvalue $\Lambda_0$ is always $0$, which is associated with the stationary state $\rho^S$ of the system satisfying ${\mathcal L} \rho^S = 0$, and all other eigenvalues have negative real parts;
This also indicates that the system has no decoherece free subspace \cite{lidar1998decoherence}.  
We assume that, reflecting the system's oscillatory dynamics, the eigenvalues with the largest non-vanishing real part (i.e., with the slowest decay rate) are given by a complex-conjugate pair.
We denote these eigenvalues as 
\begin{align}
\Lambda_1 = \mu_q - i \Omega_q, 
\quad
\overline{\Lambda_1} = \mu_q + i \Omega_q,
\end{align}
where $| \mu_q |$ $(\mu_q < 0)$ is the decay rate and $\Omega_q = \mbox{Im}\ \overline{\Lambda_1} $ gives the fundamental frequency of the oscillation.
As in the case of the stochastic oscillatory systems in Sec.~II, the oscillatory property of the system is embodied in this assumption.

\subsection{Phase space representation}

The density operator $\rho$ can also be represented by using quasiprobability distributions in the phase space such as the $P$, $Q$, and Wigner distributions
~\cite{carmichael2007statistical, gardiner1991quantum,  cahill1969density}.
We use the $P$ representation and express $\rho$ as
\begin{align}
	\rho = \int p({\bm \alpha}) | \alpha \rangle \langle \alpha |  d {\bm \alpha},
	\label{prepresentation}
\end{align}
where $| \alpha \rangle$ is a coherent state specified by 
a complex value $\alpha \in \mathbb{C}$, or equivalently by a complex vector $\bm{\alpha} = (\alpha, \overline{\alpha})^{T} \in \mathbb{C}^{2}$, $p({\bm \alpha}) : {\mathbb C}^2  \to {\mathbb R}$ is a quasiprobability 
distribution of ${\bm \alpha}$, $d{\bm \alpha} = d\alpha d\overline{\alpha}$, 
and the integral is taken over $\mathbb{C}$.
Defining the $P$ representation of an observable $F$ as
\begin{align}
	f({\bm \alpha}) = \langle \alpha | F | \alpha \rangle,
\end{align}
where $F$ is expressed in the normal order~\cite{carmichael2007statistical, gardiner1991quantum, cahill1969density}, 
the expectation of $F$ is expressed as
\begin{align}
	\langle F \rangle = \mbox{Tr} (\rho F) = \int p({\bm \alpha}) f({\bm \alpha}) d{\bm \alpha} = \mean{p(\bm{\alpha}), f(\bm{\alpha})}_{\bm{\alpha}}.
\end{align}
Here, we defined the $L^2$ inner product $\mean{g(\bm{\alpha}), h(\bm{\alpha})}_{\bm{\alpha}} = \int \overline{ g(\bm{\alpha}) } h(\bm{\alpha}) d \bm{\alpha}$
of two functions $g(\bm{\alpha})$, $h(\bm{\alpha}) : {\mathbb C}^2 \to {\mathbb C}$, where the integral is taken over the complex plane.

In the Schr\"odinger picture, the time evolution of $p_t({\bm \alpha})$ (dependence on $t$ is explicitly denoted) corresponding to the master equation~(\ref{eq:me}) is describied by a partial differential equation
\begin{align}
	\frac{\partial}{\partial t} p_t({\bm \alpha}) = {L}_{\bm \alpha} p_t({\bm \alpha}),
	\label{eq:qpde}
\end{align}
where the differential operator ${L}_{\bm \alpha}$ is related to the Liouville operator ${\mathcal L}$ in Eq.~(\ref{eq:me}) via
\begin{align*}
	{\mathcal L}\rho_t = \int {L}_{\bm \alpha} p_t({\bm \alpha}) | \alpha \rangle \langle \alpha | d{\bm \alpha}
\end{align*}
and can be explicitly calculated from Eq.~(\ref{eq:me})
by using the standard calculus for the phase-space representation
~\cite{carmichael2007statistical, gardiner1991quantum,  cahill1969density}.

The corresponding evolution of the $P$ representation $f_t({\bm \alpha})$ of the observable $F_t$ in the Heisenberg picture is given by
\begin{align}
	\frac{\partial}{\partial t} f_t({\bm \alpha}) = L_{\bm \alpha}^* f_t({\bm \alpha}),
\end{align}
where the differential operator $L_{\bm \alpha}^*$ is the adjoint of $L_{\bm \alpha}$, i.e.,
\begin{align}
	\mean{ L_{\bm \alpha} g(\bm{\alpha}), h(\bm{\alpha})}_{\bm{\alpha}} =\mean {g(\bm{\alpha}), L^*_{\bm \alpha} h(\bm{\alpha})}_{\bm{\alpha}},
\end{align}
and satisfies
\begin{align}
	L_{\bm \alpha}^* f_t({\bm \alpha}) = \langle \alpha | {\mathcal L}^* F_t | \alpha \rangle.
\end{align}
Thus, $L_{\bm \alpha}^*$ is the generator of the Koopman operator in the $P$ representation describing the evolution of $f(\bm{\alpha})$, which corresponds to the adjoint Liouville operator ${\mathcal L}^*$ in Eq.~(\ref{eq:bme}).

Corresponding to the Liouville operators ${\mathcal L}$ and ${\mathcal L}^*$, the differential operators $L_{\bm \alpha}$ and $L^*_{\bm \alpha}$ also possess a biorthogonal eigensystem $\{ \Lambda_{k}, {u}_{k}({\bm \alpha}), {v}_{k}({\bm \alpha}) \}_{k = 0, 1, 2, ...}$
of eigenvalue $\Lambda_k$ and eigenfunctions $u_k$ and $v_k$, satisfying
\begin{align}
	&L_{\bm \alpha} {u}_{k} = \Lambda_{k} {u}_{k},
	\quad
	L_{\bm \alpha}^* {v}_{k} = \overline{\Lambda_{k}} {v}_{k},
	\quad
\mean{ {u}_{k}, {v}_{l}}_{\bm \alpha} 
	= \delta_{kl}, 
\end{align}
which has one-to-one correspondence with Eq.~(\ref{eq:eigentriplet1}).
The eigenvalues $\{ \Lambda_{k} \}_{k=0,1,2,...}$ are the same as those of ${\mathcal L}$, and the eigenfunctions ${u}_{k}$ and ${v}_{k}$ of $L_{\bm \alpha}$ are related to
the eigenoperators $U_{k}$ and $V_{k}$ of ${\mathcal L}$ via
\begin{align}
	U_{k} = 
	\int {u}_{k}({\bm \alpha}) 
	| \alpha \rangle \langle \alpha |
	d{\bm \alpha},
	\quad
	{v}_{k}({\bm \alpha}) = 
	\langle \alpha | V_{k} | \alpha \rangle,
\end{align}
which follow from
\begin{align}
{\mathcal L} U_{k}
&= 
\int {u}_{k}({\bm \alpha}) \left\{ {\mathcal L} 
| \alpha \rangle \langle \alpha | \right\} d{\bm \alpha}
= 
\int \left\{ {L}_{\bm \alpha} {u}_{k}({\bm \alpha}) \right\}
| \alpha \rangle \langle \alpha | d{\bm \alpha}
= 
\int \Lambda_k {u}_{k}({\bm \alpha}) 
| \alpha \rangle \langle \alpha | d{\bm \alpha}
=
\Lambda_{k} U_{k}
\end{align}
and
\begin{align}
{L}_{\bm \alpha}^* {v}_{k} 
&= {L}_{\bm \alpha}^* \langle \alpha | V_{k} | \alpha \rangle
= 
\langle \alpha | {\mathcal L}^* V_{k} | \alpha \rangle
= \overline{\Lambda_{k}} \langle \alpha | V_{k} | \alpha \rangle
=
\overline{\Lambda_{k}} {v}_{k}.
\end{align}

\subsection{Quantum asymptotic phase}

Generalizing the definition for classical stochastic oscillatory systems in Sec.~II, we here propose a definition of the quantum asymptotic phase.
We note that, in quantum systems, the system state is given by the density operator $\rho$ and individual trajectories as in the classical stochastic systems cannot be considered. 

First, we define the quantum asymptotic phase $\Phi_q({\bm \alpha})$ of the coherent state $\bm{\bm \alpha}$ in the $P$ representation.
Considering the definition Eq.~(\ref{eq:TLphase}) of the asymptotic phase in terms of the eigenfunction $Q_1({\bm X})$ of the backward Fokker-Planck operator $L^*_{\bm X}$ in the classical stochastic case, we define the quantum asymptotic phase $\Phi_q(\bm{\alpha})$ as the argument
of the complex conjugate of the eigenfunction $v_1({\bm \alpha})$ in the $P$ representation associated with the principal eigenvalue $\overline{\Lambda_1}$ as
\begin{align}
	\label{eq:qiso}
	\Phi_q({\bm \alpha}) = \arg { {v}_1({\bm \alpha}) } = \arg \langle \alpha | { V_1 } | \alpha \rangle.
\end{align}
Next, considering that the general quantum state $\rho$ is represented as a superposition of coherent states with the weight $p({\bm \alpha})$, Eq.~(\ref{prepresentation}), we define the asymptotic phase of $\rho$ as 
\begin{align}
\Phi_q(\rho) 
= \arg \langle  p({\bm \alpha}), { v_1({\bm \alpha}) } \rangle_{\bm \alpha} 
= \arg \langle \rho, { V_1 } \rangle_{tr}.
\end{align}

It can be shown that the asymptotic phase $\Phi_q(\rho)$ evolves with a constant frequency $\Omega_q$ as the quantum state $\rho$ evolves according to the master equation~(\ref{eq:me}). 
Defining 
\begin{align}
\Psi_q(\rho_t) 
= \langle  p_t({\bm \alpha}), { v_1({\bm \alpha}) } \rangle_{\bm \alpha} 
=  \langle \rho_t, { V_1 } \rangle_{tr},
\end{align}
where the dependence on time $t$ is explicitly denoted, we have
\begin{align}
\left. \frac{d}{dt} \Psi_q(\rho_t) \right|_{t=t_0}
&= \left \langle \left. \frac{\partial p_t({\bm \alpha})}{\partial t} \right|_{t=t_0}, \ { v_1({\bm \alpha}) } \right \rangle_{\bm \alpha}
= \langle  L_{\bm \alpha} p_{t_0}({\bm \alpha}), \ { v_1({\bm \alpha}) } \rangle_{\bm \alpha} 
\cr
&=
\langle  p_{t_0}({\bm \alpha}), \ L^*_{\bm \alpha} { v_1({\bm \alpha}) } \rangle_{\bm \alpha} 
=
\langle   p_{t_0}({\bm \alpha}), \overline{\Lambda_1} { v_1({\bm \alpha}) } \rangle_{\bm \alpha} 
=
\overline{\Lambda_1} \Psi_q(\rho_{t_0}),
\end{align}
or, equivalently,
\begin{align}
\left. \frac{d}{dt} \Psi_q(\rho_t) \right|_{t=t_0} 
&= \langle \dot{\rho}_t|_{t=t_0}, V_1 \rangle_{tr}
= \langle {\mathcal L} {\rho}_{t_0}, V_1 \rangle_{tr}
= \langle  \rho_{t_0}, {\mathcal L}^* V_1 \rangle_{tr}
\cr
&= \langle \rho_{t_0}, \overline{\Lambda_1} V_1 \rangle_{tr}
= \overline{\Lambda_1} \langle  \rho_{t_0}, V_1 \rangle_{tr}
= \overline{\Lambda_1} \Psi_q(\rho_{t_0}).
\end{align}
Integrating by time, we obtain
\begin{align}
\Psi_q(\rho_t) = \exp( \overline{\Lambda_1} t ) \Psi_q(\rho_0)
=
\exp[ (\mu_q + i \Omega_q ) t ] \Psi_q(\rho_0),
\end{align}
where $\rho_0$ is the initial state at $t=0$, and hence the asymptotic phase is given by
\begin{align}
\Phi_q(\rho_t) = \arg \Psi_q(\rho_t) 
= \Omega_q t + \arg \Psi_q(\rho_0).
\end{align}
Differentiating by $t$, we obtain
\begin{align}
\frac{d}{dt} \Phi_q(\rho_t) = \Omega_q,
\end{align}
namely, the asymptotic phase $\Phi_q(\rho_t)$ increases with a constant frequency $\Omega_q$ with the evolution of the quantum state $\rho_t$.

Thus, by using the eigenfunction $v_1({\bm \alpha})$ of the adjoint linear operator $L^*_{\bm \alpha}$ or equivalently the eigenoperator $V_1$ of the adjoint operator ${\mathcal L}^*$ associated with the eigenvalue $\overline{\Lambda_1}$, we can define the asymptotic phase $\Phi_q(\rho)$ of the quantum state $\rho$.
The quantum master equation~(\ref{eq:me}), adjoint Liouville operator ${\mathcal L}^*$ (or adjoint differential operator $L^*_{\bm \alpha}$ in the $P$ representation), and eigenoperator $V_1$ (or the eigenfunction $v_1({\bm \alpha})$ in the $P$ representation) with the eigenvalue $\overline{\Lambda_1}$ correspond to the forward Fokker-Planck equation~(\ref{forward}), backward Fokker-Planck operator $L_{\bm X}^*$, and eigenfunction ${ Q_1 (\bm{X}) }$
with the eigenvalue $\overline{ \lambda_1}$ in the classical stochastic system discussed in Sec.~II, respectively. 

We stress, however, that the system state is generally represented by the density operator $\rho$ or quasiprobability distribution $p({\bm \alpha})$ and individual trajectories cannot be considered in the quantum case.  
As in the classical stochastic case, we may also choose the eigenvalue $\Lambda_1$, eigenfunction $\overline{v_1 (\bm{\alpha})}$, and eigenoperator $V_1^{\dag}$ to define the quantum asymptotic phase, which reverses its direction of increase.

\section{Example: quantum van der Pol oscillator}

In this section, using the quantum van der Pol model with quantum Kerr effect as an example, we illustrate the validity of the quantum asymptotic phase defined in Sec.~III.
We also analyze a damped harmonic oscillator in Appendix D.

\subsection{Quantum van der Pol model with quantum Kerr effect}

As an example of quantum oscillatory systems, we consider the quantum van der Pol model with quantum Kerr effect. The system's density operator $\rho$ obeys the master equation
\begin{align}
	\label{eq:qvdp_me}
	\dot{\rho} 
	= {\mathcal L} \rho
	= - i \left[ 
	H
	,\rho\right]
	+ \gamma_{1} \mathcal{D}[a^{\dag}]\rho + \gamma_{2}\mathcal{D}[a^{2}]\rho,
\end{align}
where $a$ and $a^\dag$ are annihilation and creation operators, $H = \omega_0 a^{\dag}a + K a^{\dag 2} a^2$ is the Hamiltonian,
$\omega_{0}$ is the frequency parameter of the oscillator, 
$K$ is the Kerr parameter,
and $\gamma_{1}$ and $\gamma_{2}$ are the decay rates for 
negative damping and nonlinear damping, respectively~\cite{kato2020semiclassical,lorch2016genuine}.
By using the standard rule of calculus~\cite{carmichael2007statistical, gardiner1991quantum, cahill1969density}, we can derive the differential operator $L_{\bm \alpha}$ in Eq.~(\ref{eq:qpde}) describing the evolution of the $P$ representation $p({\bm \alpha})$ of $\rho$ from the Liouville operator ${\mathcal L}$, which consists of third- and higher-order derivative with respect to $\alpha$ \cite{lorch2016genuine}.

To evaluate the fundamental frequency $\Omega_q$ and the asymptotic phase $\Phi_q(\rho)$ of the system, we numerically calculate the eigenvalues and eigenfunctions of the Liouville and adjoint Liouville operators ${\mathcal L}$ and ${\mathcal L}^*$.
To this end, we approximately truncate the number representation of the density operator as a large $N \times N$ matrix and map it to a $N^2$-dimensional vector of the double-ket notation \cite{albert2018lindbladians}.
We can then approximately represent the Liouville operators ${\mathcal L}$ and ${\mathcal L}^*$ by 
$N^2 \times N^2$ matrices and calculate their eigensystem to obtain the asymptotic phase in Eq.~(\ref{eq:qiso}).

\subsection{Semiclassical regime}

We first consider the semiclassical regime where $\gamma_{2}$ and $K$ are sufficiently small. In this case, as explained in Appendix B, we can approximate Eq.~(\ref{eq:qpde}) by a Fokker-Planck equation for $p({\bm \alpha})$, namely, the system state is approximately equivalent to a classical stochastic system.
Furthermore, in the classical limit where the quantum noise vanishes, the system is described by a single complex variable $\alpha \in {\mathbb C}$ obeying a deterministic ordinary differential equation
\begin{align}
	\dot{\alpha} = \left( \frac{\gamma_1}{2} - i \omega_0 \right) \alpha 
	- (\gamma_{2}  + 2  K  i ) \overline{\alpha} \alpha^{2}.
	\label{eq:qvdp_ldvm}
\end{align}
This equation represents the Stuart-Landau oscillator (normal form of the supercritical Hopf bifurcation)~\cite{kuramoto1984chemical} and possesses a stable limit-cycle solution
${\alpha}_0(\phi) = R e^{i \phi}$,
which is represented as a function of the phase $\phi = \Omega_c t +const.$ with a
natural frequency $\Omega_c = -\omega_0 - K \gamma_1/\gamma_2$
and radius 
$R = \sqrt{{\gamma_1} / {2 \gamma_{2} }}$.
The basin $B$ of this limit cycle 
is the whole complex plane except the origin. 
The classical asymptotic phase $\Phi_c$ of this system is given by~\cite{kato2019semiclassical}
\begin{align}
	\label{eq:qvdp_iso}
	\Phi_c({\bm \alpha}) = \mbox{arg}\ \alpha - \frac{2 K}{\gamma_{2}} \ln \frac{ | \alpha | }{R} + const.
\end{align}
and satisfies $\dot{\Phi}_c({\bm \alpha}) = \Omega_c$ as ${\alpha}$ evolves in $B$ under Eq.~(\ref{eq:qvdp_ldvm})~\cite{nakao2016phase}.
As stated in Sec.~II, this $\Phi_c({\bm \alpha})$ is the argument of the Koopman eigenfunction $\Psi_c({\bm \alpha})$ of the system associated with the eigenvalue $i \Omega_c$.
In Ref.~\cite{kato2019semiclassical}, we used this $\Phi_c$ for the phase-reduction analysis of quantum synchronization in the semiclassical regime with weak quantum noise.
It is expected that the quantum asymptotic phase $\Phi_q( \bm \alpha)$
of the coherent state $\bm{\alpha}$
is close to the classical asymptotic phase $\Phi_c( \bm \alpha)$ when the quantum noise is sufficiently small.

\begin{figure} [!t]
	\begin{center}
		\includegraphics[width=0.9\hsize,keepaspectratio]{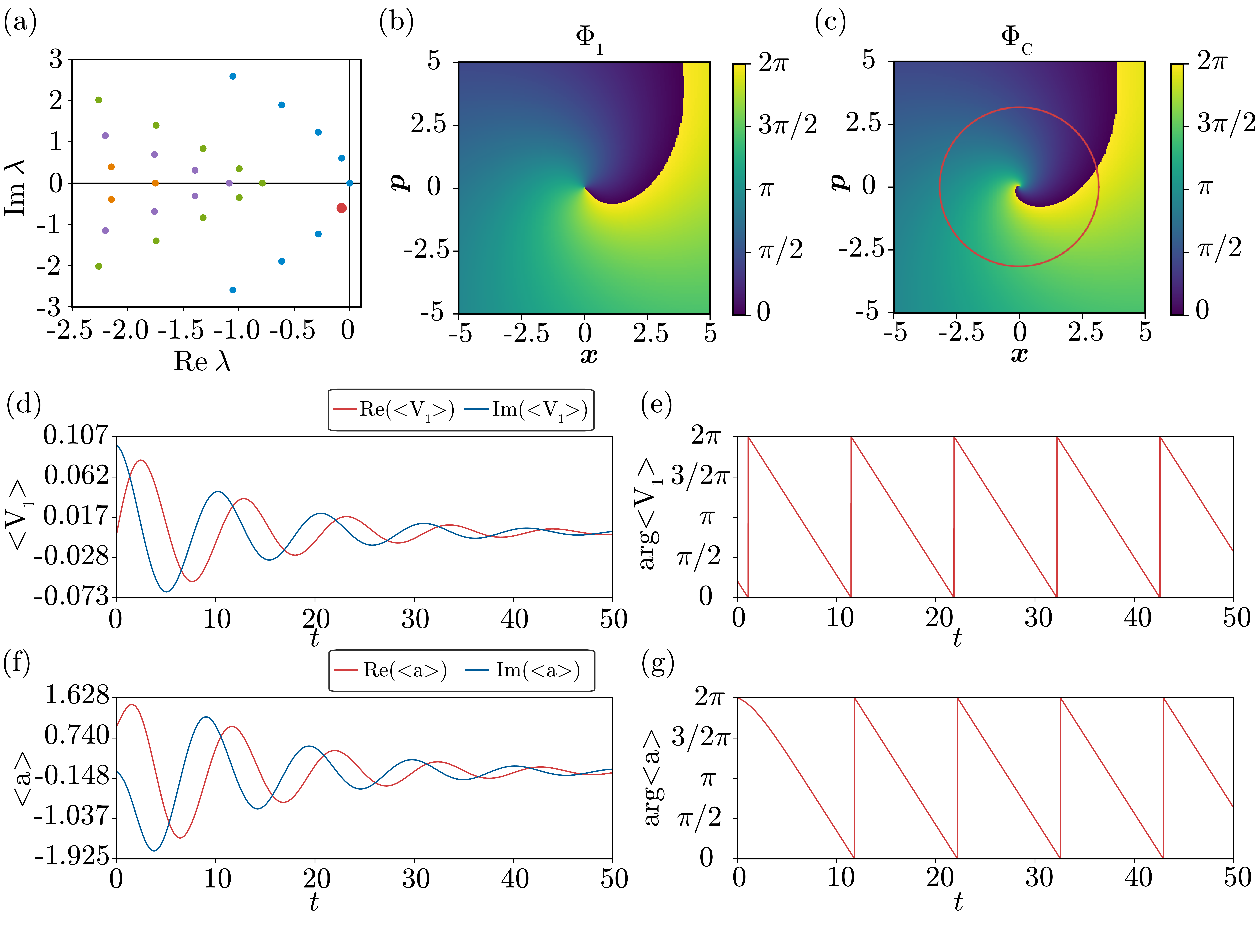}
	\end{center}
	\caption{
		Quantum asymptotic phase in the semiclassical regime. 
		The parameters are $\gamma_1=1$ and $(\omega_0, \gamma_{2}, K)/\gamma_{1} = ({0.1}, 0.05, 0.025)$.
		(a) Eigenvalues of ${\mathcal L}_0$ near the imaginary axis. The red dot represents the principal eigenvalue 
		{$\overline{\Lambda_1}$}
		with the the slowest decay rate.
		(b) Quantum asymptotic phase $\Phi_q$ with $\Omega_q = 
		{-0.605}$.
		(c) Classical asymptotic phase $\Phi_c$ with $\Omega_c = { -0.6}$.
		(d-g) Evolution of the expectation values of $V_1$ and $a$ and their arguments from a pure coherent state:
		(d) $\Psi(\rho_t) = \mean{V_1}_t$, (e) $\Phi(\rho_t) = \arg \mean{V_1}_t$, (f) $\mean{a}_t$, and (g) $\arg \mean{a}_t$.
		In (a), individual branches of eigenvalues are shown with different colors.
		In (b), (c),  $(x,p)=(2.5, 0)$ is chosen as the phase origin.
		In (c), the red-thin line represents the limit cycle in the classical limit.
		}
	\label{fig_1}
\end{figure}

Figure~\ref{fig_1}(a) shows the eigenvalues of ${\mathcal L}$ near the imaginary axis obtained numerically, where the principal eigenvalue 
$\overline{\Lambda_1} = \mu_q + i \Omega_q$ is shown by a red dot ($\mu_q < 0$),
and
Figs.~\ref{fig_1}(b) and~\ref{fig_1}(c) compare the quantum-mechanical phase $\Phi_q( \bm \alpha)$ 
with the corresponding classical phase $\Phi_c({\bm \alpha})$.
Here, we adopt a negative value for $\Omega_q$ so that the resulting phase  $\Phi_q$ increases  
in the counterclockwise direction from $0$ to $2\pi$ on the complex plane, i.e., $\Phi_q$ satisfies $\oint_c \nabla \Phi_q({\bm x}) \cdot d{\bm x} = 2\pi$ where ${\bm x} = (x, p) = ( \mbox{Re}~\alpha, \mbox{Im}~\alpha )$ 
and $C$ is a circle around $0$.

As the quantum noise is small, the quantum frequency $\Omega_q$ and the asymptotic phase $\Phi_q(\bm{\alpha})$ 
obtained numerically from ${\mathcal L}$ are close to the classical frequency $\Omega_c$ and asymptotic phase $\Phi_c({\bm \alpha})$, respectively.
The difference between $\Omega_q$ and $\Omega_c$ arises from the small quantum noise;
in the limit of vanishing quantum noise, the eigenfunction ${v}_1({\bm \alpha})$ of $L_{\bm \alpha}$ in the $P$ representation coincides with the Koopman eigenfunction of the deterministic system Eq.~(\ref{eq:qvdp_ldvm}) with the eigenvalue $i \Omega_c$ and therefore $\Phi_q$ reproduces the classical phase $\Phi_c$ (see Appendix  C).
Here, we note that the principal eigenvalues $i \Omega_q$ and $i \Omega_c$ are well separated from other branches of eigenvalues with faster decay rates and the corresponding oscillatory modes become quickly dominant.

To demonstrate that the present definition of the quantum asymptotic phase yields appropriate values, we consider free oscillatory relaxation of $\rho_t$ from a coherent initial state $\rho^{\alpha_0} = \ket{\alpha_0}\bra{\alpha_0}$ with $\alpha_0=1$ at $t=0$ and measure $\Phi_q(\rho_t)$.
For comparison, we also measure the argument $\mbox{arg}\ \langle a \rangle_t$ of the expectation $\langle a \rangle_t = \langle \rho_t, a \rangle_{tr}$ of the annihilation operator $a$, which simply gives the polar angle of $\langle a \rangle_t$ on the complex plane.
We note that, although the system state $\rho$ starts from a pure coherent state, it soon becomes a mixed state due to the coupling with the reservoirs and eventually relaxes
to the stationary state $\rho^S$.

Figures~\ref{fig_1}(d) and (e) plot the evolution of the expectation $\Psi_q(\rho_t) = \langle \rho_t, V_1 \rangle_{tr}$ of the eigenoperator $V_1$ and the quantum asymptotic phase $\Phi_q(\rho_t) = \arg \Psi_q(\rho_t)$, respectively,
and
Figs.~\ref{fig_1}(f) and (g) show the evolution of the expectation $\langle a \rangle$ and its polar angle $\arg \langle a \rangle$, respectively.
The asymptotic phase $\Phi_q(\rho_t)$ increases with a constant frequency $\Omega_q$ and appropriately yields isochronous phase values.
In contrast, the polar angle $\arg \langle a \rangle$ does not increase constantly with time, in particular in the transient process before $t=10$, as shown in Fig.~\ref{fig_1}(g); as the limit cycle in the classical limit is rotationally symmetric in this model, the polar angle also yields almost constantly increasing phase values after relaxation. 

Thus, the quantum asymptotic phase $\Phi_q$ increases with a constant frequency $\Omega_q$ in the semiclassical regime of a quantum van der Pol model with the quantum Kerr effect.

\subsection{Strong quantum regime}

\begin{figure} [!t]
	\begin{center}
		\includegraphics[width=0.9\hsize,clip]{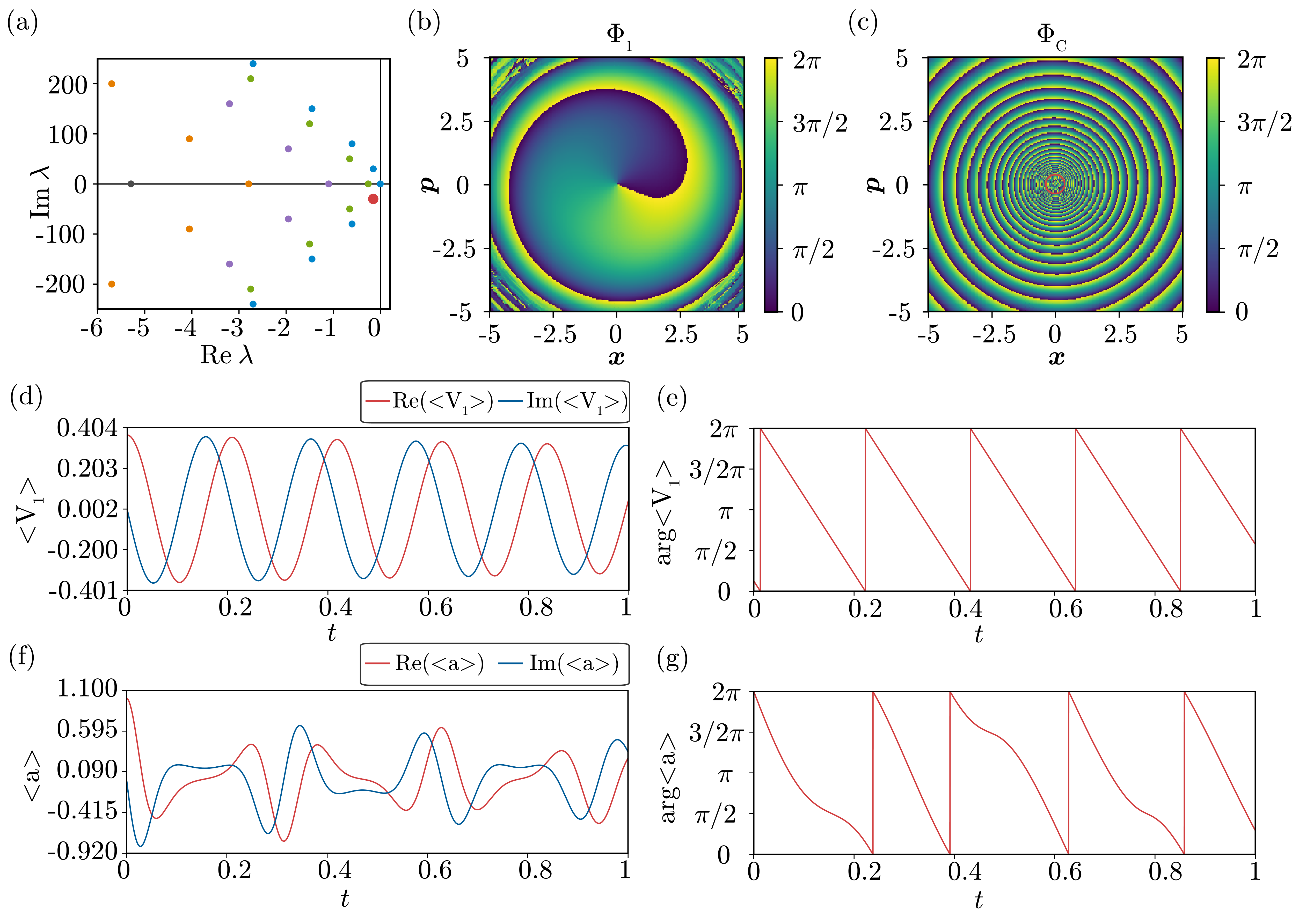}
		\caption{
		Quantum asymptotic phase in the strong quantum regime. 
		The parameters are $\gamma_1=0.1$ and $(\omega_0, \gamma_{2}, K)/\gamma_{1} = (300, 4, 100)$.
		(a) Eigenvalues of ${\mathcal L}$ near the imaginary axis. The red dot represents the principal eigenvalue
		{$\overline{\Lambda_1}$} with the the slowest decay rate.
		(b) Quantum asymptotic phase $\Phi_q$ with $\Omega_q = -30$.
		(c) Classical asymptotic phase $\Phi_c$ with $\Omega_c = -32.5$.
		(d-g) Evolution of the expectation values of $V_1$ and $a$ and their arguments from a pure coherent state:
		(d) $\Psi(\rho_t) = \mean{V_1}_t$, (e) $\Phi(\rho_t) = \arg \mean{V_1}_t$, (f) $\mean{a}_t$, and (g) $\arg \mean{a}_t$.
		In (a), individual branches of eigenvalues are shown with different colors.
		In (b), (c),  $(x,p)=(2.5, 0)$ is chosen as the phase origin.
		In (c), the red-thin line represents the limit cycle in the classical limit.
		}
		\label{fig_2}
	\end{center}
\end{figure}

Next, we consider a strong quantum regime with relatively large $\gamma_{2}$ and $K$, where only a small number of energy states participates in the system dynamics and the semiclassical description is not valid.
The eigenvalues of ${\mathcal L}$ obtained numerically are shown in Fig.~\ref{fig_2}(a).
Figures~\ref{fig_2}(b) and~\ref{fig_2}(c) show the quantum-mechanical phase  $\Phi_q$ and the corresponding classical phase $\Phi_c$.
Because the system is in the strong quantum regime,
$\Phi_c$ is distinctly different from $\Phi_q$ and the classical frequency $\Omega_c$ also differs largely from the true quantum frequency $\Omega_q$.
Here, we again note that the principal eigenvalues have much smaller (less than half of the second largest) decay rate than the eigenvalues in the other branches.

We consider free oscillatory relaxation of $\rho$ from a coherent initial state $\rho =  \ket{\alpha_0}\bra{\alpha_0}$ with $\alpha_0=1$ at $t=0$ and measure the evolution of the asymptotic phase $\Phi_q(\rho_t) = \langle V_1 \rangle_t$ of the system state $\rho_t$.
For comparison, we also measure the polar angle $\mbox{arg} \langle a \rangle_t$ of $\langle a \rangle_t$.

Figures~\ref{fig_2}(d) and (e) plot the evolution of the expectation $\Psi_q(\rho_t) = \langle \rho_t, V_1 \rangle_{tr}$ of the eigenoperator $V_1$ and the quantum asymptotic phase $\Phi_q(\rho_t) = \arg \Psi_q(\rho_t)$, respectively,
and
Figs.~\ref{fig_2}(f) and (g) show the evolution of the expectation $\langle a \rangle$ and its polar angle $\arg \langle a \rangle$, respectively.
As expected, the asymptotic phase $\Phi_q(\rho_t)$ appropriately gives constantly varying phase values with the frequency $\Omega_q$.
In contrast, the polar angle $\arg \langle a \rangle_t$ does not vary constantly with time and is not isochronous.
This is because the transition between relatively a small number of energy levels takes part in the system dynamics and the discreteness of the energy spectra can {play important roles} in this strong quantum regime.

Thus, the quantum asymptotic phase $\Phi_q$ gives appropriate phase values that increase with a constant frequency $\Omega_q$ even in the strong quantum regime of a quantum van der Pol model with quantum Kerr effect, even though the strong quantum effect strongly alters the dynamics of the system from the classical limit.

\subsection{Fundamental difference between the classical and quantum systems}

Although we introduced the definition of the quantum asymptotic phase $\Phi_q$ by analogy with the classical deterministic phase $\Phi_c$ and classical stochastic asymptotic phase $\Phi_s$, a fundamental difference exists between the quantum and classical cases.
Specifically, in the quantum case, the system state is described by the density operator $\rho$ and the asymptotic phase $\Phi_q$ assigns a phase value on each $\rho$, while in the classical case,
the phase function $\Phi_c$ or $\Phi_s$ assigns a phase value on each individual state ${\bm X}$,
although the probability density function $p({\bm X})$ is used in defining the stochastic asymptotic phase $\Phi_s$.
This difference arises because the system state can be described by a SDE representing a single stochastic trajectory of the noisy oscillator in the classical stochastic case, whereas the system state can only be characterized by the density operator $\rho$ representing the
statistical state of the system in the quantum case.

Though we have taken the viewpoint that the asymptotic phase is defined for a single stochastic trajectory in the classical stochastic case in this study, we may also consider that the asymptotic phase is defined for the probability density $p({\bm X})$ rather than for the state ${\bm X}$ in the classical stochastic case.
Then, the quantum asymptotic phase $\Phi_q$ as a function of the density matrix $\rho$ corresponds to the stochastic asymptotic phase $\Phi_s(p({\bm X})) = \arg \int p({\bm X}) Q_1({\bm X}) d{\bm X}$ as a function of the probability density $p({\bm X})$ in the classical stochastic case in Sec. II.
We regarded this quantity as the exponential average of the asymptotic phase values defined for individual states in Eq.~(\ref{expaveg}).

In the present approach, it is always possible to formally introduce a 'phase function' for any oscillatory system as long as it has the decaying mode with a non-zero imaginary part by using the associated eigenoperator. However,  this phase function does not necessarily play the role of a quantum asymptotic phase unless the system exhibits limit-cycle oscillations and synchronization. As an example, in Appendix D, we introduce a phase function of a damped quantum harmonic oscillator, which cannot be considered the quantum asymptotic phase since the system does not exhibit synchronization. Only when the system is nonlinear and exhibits synchronization, the asymptotic phase captures the synchronization dynamics of the system.

The asymptotic phase is the basis for developing phase reduction theory for classical nonlinear oscillators. For quantum oscillatory systems, however, even if we can introduce the asymptotic phase as described in this study, it does not necessarily mean that we can develop the phase reduction theory. This is because the quantum state $\rho$ may not be appropriately localized in the phase-space representation and hence it may not be reconstructed from the phase value even approximately. Still, as we showed for the case of the quantum vdP oscillator in the semiclassical regime~\cite{kato2019semiclassical}, we may approximately describe the dynamics of the quantum state by using the reduced phase equation and perform a detailed synchronization analysis in appropriate physical situations. It should be noted that we also encounter a similar problem in developing phase reduction theory for classical oscillators under strong noise.
We also note that, even if we cannot derive a reduced phase description for quantum nonlinear oscillators, the asymptotic phase can be used to characterize the peculiar properties of quantum synchronization~\cite{kato2021koopman2}.
%
\section{Summary}

In this study, we proposed a definition of the asymptotic phase for quantum oscillatory systems by generalizing the asymptotic phase for classical stochastic oscillatory system proposed by Thomas and Lindner~\cite{thomas2014asymptotic} from the Koopman operator viewpoint~\cite{kato2021asymptotic}. 
The proposed asymptotic phase is defined by using the eigenoperator of the adjoint Liouville operator describing the evolution of the quantum-mechanical observable, in close analogy to the asymptotic phase for classical limit-cycle oscillators that can be interpreted as the argument of the Koopman eigenfunction associated with the fundamental frequency.
By using the quantum van der Pol model with quantum Kerr effect as an example, we demonstrated that the proposed asymptotic phase appropriately yields isochronous phase values even in the strong quantum regime where the semiclassical approximation is not valid.

Though quantum synchronization has attracted much attention recently,
compared with classical synchronization,
systematic analysis of the quantum synchronization has been restricted,
partly due to the lack of the clear definition of the phase.
The proposed definition of the asymptotic phase valid in strong quantum regimes may be used for systematic and quantitative analysis of synchronization phenomena in quantum nonlinear oscillators~\cite{kato2021koopman2}.
Moreover, we may be able to develop a phase reduction theory for strongly quantum nonlinear oscillators by using the proposed definition, which would allow us to reduce the system dynamics to a simple phase equation and facilitates detailed analysis, control, and optimization of quantum nonlinear oscillators.
It will also be interesting to extend the definition of the amplitude functions for classical stochastic oscillators  \cite{perez2021isostables, kato2021asymptotic} to strongly quantum oscillatory systems on the basis of the Koopman operator theory.

\begin{acknowledgements}
	The authors thank anonymous reviewers for their enlightning comments.
	Numerical simulations have been performed by using QuTiP numerical
	toolbox~\cite{johansson2012qutip,johansson2013qutip}. 
	This research was funded by JSPS KAKENHI JP17H03279, JP18H03287, JPJSBP120202201, JP20J13778, 
	JP22K14274, JP22K11919, JP22H00516 and JST CREST JP-MJCR1913.
\end{acknowledgements}

\subsection*{Data availability}
The data that supports the findings of this study are available within the article.

\appendix

\section{Koopman operator for classical stochastic processes}

In this section, we briefly summarize the definition of the Koopman operator for classical stochastic processes described by the Ito SDE~(\ref{sde}).
The evolution of an observable $g : {\mathbb R}^N \to {\mathbb C}$ for this Ito diffusion process from time $t_0$ to ${t}+{t_0}$ is expressed as~\cite{mezic2005spectral,oksendal2013stochastic,vcrnjaric2020koopman}
\begin{align}
g_{t+t_0}({\bm X}) = (U^{t} g_{t_0})({\bm X}),
\end{align}
where $U^{t}$ (${t} \geq 0$) is the stochastic Koopman operator defined as 
\begin{align}
(U^{{t}} g) ({\bm Y}) 
&= {\mathbb E}^{\bm Y}[ g({\bm X}({t}+{t_0})) ] 
\cr
&= \int g({\bm X}) p({\bm X}, {t}+{t_0} | {\bm Y}, t_0 ) d{\bm X} = \langle p({\bm X}, {t}+{t_0} | {\bm Y}, t_0 ), g({\bm X}) \rangle_{\bm X}.
\end{align}
Here, ${\mathbb E}^{\bm Y}[ \cdot ]$ represents the expectation over the stochastic realizations of ${\bm X}({t})$ started from ${\bm X}(t_0) = {\bm Y}$ at $t=t_0$,
$p(\bm {X}, t| \bm {Y}, s)$ ($t \geq s$) is the transition probability density, and $\langle \cdot, \cdot \rangle_{\bm X}$ represents the inner product defined in Sec. II.
Defining a time evolution operator $S^{t} = e^{{t} L_{\bm X}}$ (${t} \geq 0$),
the evolution of the probability density  $p({\bm X})$ is expressed as
\begin{align}
p_{{t}+{t_0}}({\bm X}) = (S^{{t}} p_{t_0})({\bm X}),
\end{align}
where the operation of $S^{t}$ on a function $f : {\mathbb R} \to {\mathbb C}$ is given by
\begin{align}
(S^{{t}} f)({\bm X}) = \int p({\bm X}, {t}+{t_0} | {\bm Y}, t_0) f({\bm Y}) d{\bm Y}.
\end{align}
The Koopman operator $U^{t}$ is the adjoint of $S^{t}$, i.e., 
\begin{align}
	\langle  f({\bm X}),\ (U^{{t}} g)({\bm X}) \rangle_{\bm X} 
	&= \int \overline{ f({\bm Y}) } \ \left[ \int g({\bm X}) p({\bm X}, {t}+{t_0} | {\bm Y}, t_0) d{\bm X} \right] d{\bm Y}
	\cr
	&= \int \overline{ \left[ \int p({\bm X}, {t}+{t_0} | {\bm Y}, t_0) f({\bm Y}) d{\bm Y} \right] } g({\bm X}) d{\bm X}
	\cr
	&= \int \overline{ (S^t f)({\bm X}) } g({\bm X}) d{\bm X}
	= \langle (S^{{t}} f)({\bm X}),\ {g({\bm X})} \rangle_{\bm X}
\end{align}
for two functions $f, g : {\mathbb R}^N \to {\mathbb C}$.
It is noted that the expectation of the observable $g$ at time ${t}+{t_0}$ can be expressed as
\begin{align}
\int p_{t_0}({\bm X}) (U^{t} g_{t_0})({\bm X}) d{\bm X} 
&= 
\langle p_{t_0}({\bm X}),\ { (U^{{t}} g_{t_0})({\bm X})} \rangle_{\bm X}
\cr
&= \langle (S^{{t}} p_{t_0})({\bm X}),\ { g_{t_0}({\bm X}) } \rangle_{\bm X}
=
\int (S^t p_{t_0})({\bm X}) g_{t_0}({\bm X}) d{\bm X}.
\end{align}

It can be shown that the infinitesimal generator of $U^{t}$ is given by the backward operator $L_{\bm X}^*$ in Eq.~(\ref{eq:Lxadj}) (see e.g. Ref.~\cite{oksendal2013stochastic} for a rigorous treatment).
From the Ito formula~\cite{gardiner2009stochastic,oksendal2013stochastic}, the SDE for a function $g: {\mathbb R}^N \to {\mathbb R}$ is given by
\begin{align}
d g({\bm X}) = \left[ {\bm A}({\bm X}) \cdot \frac{\partial}{\partial {\bm X}} g({\bm X}) + \frac{1}{2} {\bm D}({\bm X}) \frac{\partial^2}{\partial {\bm X}^2} g({\bm X}) \right] dt + {\bm B}({\bm X}) \frac{\partial}{\partial {\bm X}} g({\bm X}) d{\bm W},
\end{align}
which gives
\begin{align}
{\mathbb E}^{\bm Y}[ g({\bm X}({t}+{t_0})) ] = \int_{t_0}^{{t}+{t_0}} \left[ {\bm A}({\bm X}) \cdot \frac{\partial}{\partial {\bm X}} g({\bm X}) + \frac{1}{2} {\bm D}({\bm X}) \frac{\partial^2}{\partial {\bm X}^2} g({\bm X}) \right] dt
=
\int_{t_0}^{t+t_0} L^*_{\bm X} g({\bm X}) dt
\end{align}
by integration. Assuming the function $g$ to be the observable $g_{t_0}$ at $t=t_0$, the infinitesimal evolution of $g_t$ at $t=t_0$ can be represented as
\begin{align}
\left. \frac{d}{d t} g_t({\bm Y}) \right|_{t=t_0}
=
\lim_{t \to +0} \frac{ (U^{t} g_{t_0})({\bm Y}) - g_{t_0}({\bm Y})}{t}
=
\lim_{t \to +0} \frac{ {\mathbb E}^{\bm Y}[ g_{t_0}({\bm X}(t+t_0)) ] - g_{t_0}({\bm Y})}{t}
=
L_{\bm X}^* g_{t_0}({\bm Y}).
\end{align}
Thus, we can express the Koopman operator as $U^{t} = e^{{t}L^*_{\bm X}}$.

From the adjoint relation for $S^{t}$ and $U^{t}$, $L_{\bm X}$ and $L_{\bm X}^*$ are also adjoint to each other, i.e.,
$\langle L_{\bm X} f({\bm X}), g({\bm X}) \rangle_{\bm{X}}
=
\langle f({\bm X}), L^*_{\bm X} g({\bm X}) \rangle_{\bm{X}}$,
and the evolution of the expectation of the observable $g$ at time $t = t_0$ can be expressed as in Eq.~(\ref{clevobs}).

\section{Quantum van der Pol oscillator with Kerr effect in the semiclassical regime}

In this section, we briefly explain the classical limit of the quantum van der Pol oscillator.
As shown in Ref.~\cite{kato2019semiclassical}, in the semiclassical regime, the linear operator $L_{\bm \alpha}$ in Eq.~(\ref{eq:qpde}) describing the evolution of the quasiprobability distribution $p({\bm \alpha})$ in the $P$ representation
of the quantum van der Pol oscillator
can be approximated by a Fokker-Planck operator
\begin{align}
	\tilde{L}_{\bm \alpha} = \Big[ - \sum_{j=1}^{2} \partial_{j} \{ A_{j}(\bm{\alpha}) \}
	+ \frac{1}{2} \sum_{j=1}^2 \sum_{k=1}^2 \partial_{j}\partial_{k} \{ D_{jk}(\bm{\alpha}) \} \Big]
	\label{eq:qfpe}
\end{align}
by neglecting the third- and higher-order derivatives, where $\partial_1 = \partial / \partial \alpha$ and $\partial_2 = \partial / \partial \bar{\alpha}$.
The drift vector $\bm{A}(\bm{\alpha}) = \left( A_1(\bm{\alpha}), A_2(\bm{\alpha}) \right) \in {\mathbb C}^2$ and the matrix $\bm{D}(\bm{\alpha}) = \left( D_{jk}(\bm{\alpha}) \right) \in {\mathbb C}^{2 \times 2}$ are given by
\begin{align}
	\label{eq:qvdp_drift}
	\hspace{-3em}
	\bm{A}(\bm{\alpha}) 
	&=
	\left( \begin{matrix}
		\left(\frac{\gamma_1 }{2} - i \omega_0 \right) \alpha 
		- (\gamma_{2}  + 2  K  i ) \overline{\alpha}  \alpha^{2}
		\\
		\left(\frac{\gamma_1}{2} + i \omega_0  \right) \overline{\alpha}    
		- (\gamma_{2}  - 2  K  i ) \alpha \overline{\alpha}^{2}  
		\\
	\end{matrix} \right),
	\\
	\label{eq:qvdp_diffusion}
	\bm{D}(\bm{\alpha}) &= 
	\left( \begin{matrix}
		-( \gamma_{2} + 2 K i) \alpha^{2}   & \gamma_1  \\
		\gamma_1 & -( \gamma_{2} - 2 K i)  \bar{\alpha}^{2} \\
	\end{matrix} \right).
\end{align}
The corresponding stochastic differential equation is thus given by
\begin{align}
	\label{eq:qvdp_ldv}
	d
	\left( \begin{matrix}
		{\alpha}  \\
		\overline{\alpha}  \\
	\end{matrix} \right)
	&=
	\left( \begin{matrix}
		\left(\frac{\gamma_1 }{2} - i \omega_0 \right) \alpha 
		- (\gamma_{2}  + 2  K  i ) \overline{\alpha}  \alpha^{2}
		\\
		\left(\frac{\gamma_1}{2} + i \omega_0  \right) \overline{\alpha}    
		- (\gamma_{2}  - 2  K  i ) \alpha \overline{\alpha}^{2}  
		\\
	\end{matrix} \right) dt
	+\bm{\beta}(\bm{\alpha})
	\left( \begin{matrix}
		dW_1
		\\
		dW_2
		\\
	\end{matrix}
	\right)
	,
\end{align}
where $W_1$ and $W_2$ are independent Wiener processes
and the matrix ${\bm \beta}({\bm \alpha})$ is given by
\begin{align}
	\label{eq:qvdp_beta}
	\bm{\beta}(\bm{\alpha}) &= 
	\begin{pmatrix}
		\sqrt{\frac{\left( \gamma_1 +  R_{11}(\bm{\alpha}) \right)}{2}} e^{i \chi(\bm{\alpha}) / 2}
		&
		-i \sqrt{\frac{\left( \gamma_1 -  R_{11}(\bm{\alpha}) \right)}{2}} e^{i \chi(\bm{\alpha}) / 2}
		\\
		\sqrt{\frac{\left( \gamma_1 +  R_{11}(\bm{\alpha}) \right)}{2}} e^{- i \chi(\bm{\alpha}) / 2}
		&
		i \sqrt{\frac{\left( \gamma_1 -  R_{11}(\bm{\alpha}) \right)}{2}} e^{- i \chi(\bm{\alpha}) / 2}
	\end{pmatrix},
\end{align}
where $R_{11}(\bm{\alpha}) e^{i \chi(\bm{\alpha})} = -(\gamma_{2} + 2 K i) \alpha^{2}$. It is noted  that the two equations for $\alpha$ and $\overline{\alpha}$ in Eq.~(\ref{eq:qvdp_ldv}) are mutually complex conjugate and represent the same dynamics.

In the classical limit, the deterministic part of Eq.~(\ref{eq:qvdp_ldv}) gives the Stuart-Landau equation for the complex variable $\alpha$ given in Eq.~(\ref{eq:qvdp_ldvm}), 
which is analytically solvable and the asymptotic phase
$\Phi_c(\bm \alpha)$ can be explicitly obtained as given 
in Eq.~(\ref{eq:qvdp_iso})~\cite{kuramoto1984chemical,nakao2016phase}.

\section{Classical limit of the quantum asymptotic phase}

In this section, we explain that the quantum asymptotic phase formally reproduces the deterministic asymptotic phase in the classical limit.
In the semiclassical regime, the linear operator $L_{\bm \alpha}$ of Eq.~(\ref{eq:qpde}) for the quasiprobability distribution $p ({\bm \alpha})$ in the $P$ representation can be approximated by a Fokker-Planck operator $\tilde{L}_{\bm \alpha}$ of the form Eq.~(\ref{eq:qfpe}).
By introducing a real vector ${\bm X} = (\mbox{Re}\ \alpha, \mbox{Im}\ \alpha)$ and the corresponding probability density function $p({\bm X})$, the Fokker-Planck operator $\tilde{L}_{\bm \alpha}$ for $p({\bm \alpha})$ can be cast into a real Fokker-Planck operator $L_{\bm X}$ for $p({\bm X})$ in Eq.~(\ref{eq:fpe2}).

In the classical limit, the quantum noise vanishes and the diffusion term in $L_{\bm X}$ disappears.
Thus, $L_{\bm X}$ formally becomes a classical Liouville operator, i.e., $L_{\bm X} \to - (\pa/\pa {\bm X}) {\bm A}({\bm X})$, and the corresponding backward Liouville operator formally becomes the infinitesimal generator of the deterministic Koopman operator, i.e., $L_{\bm X}^* \to A = {\bm A}({\bm X}) \cdot ({\partial}/{\partial {\bm X}}) = {\bm A}({\bm X}) \cdot \nabla$.
Also, the decay rate $\mu_q$ approaches $0$ and the eigenvalue $\Lambda_q$ approaches $i \Omega_c$ where $\Omega_c$ is the frequency of the limiting classical deterministic system.
As discussed in Sec.~II~A, the classical asymptotic phase $\Phi_c$ is obtained as the argument of the eigenfunction $\Psi_c$ of $A$ associated with the eigenvalue $\overline{\Lambda_1} = i \Omega_q$.
Thus, the quantum asymptotic phase $\Phi_q$ formally reproduces the deterministic asymptotic phase $\Phi_c$ in the classical limit with vanishing quantum noise.

\begin{figure} [htbp]
	\begin{center}
		\includegraphics[width=0.7\hsize,clip]{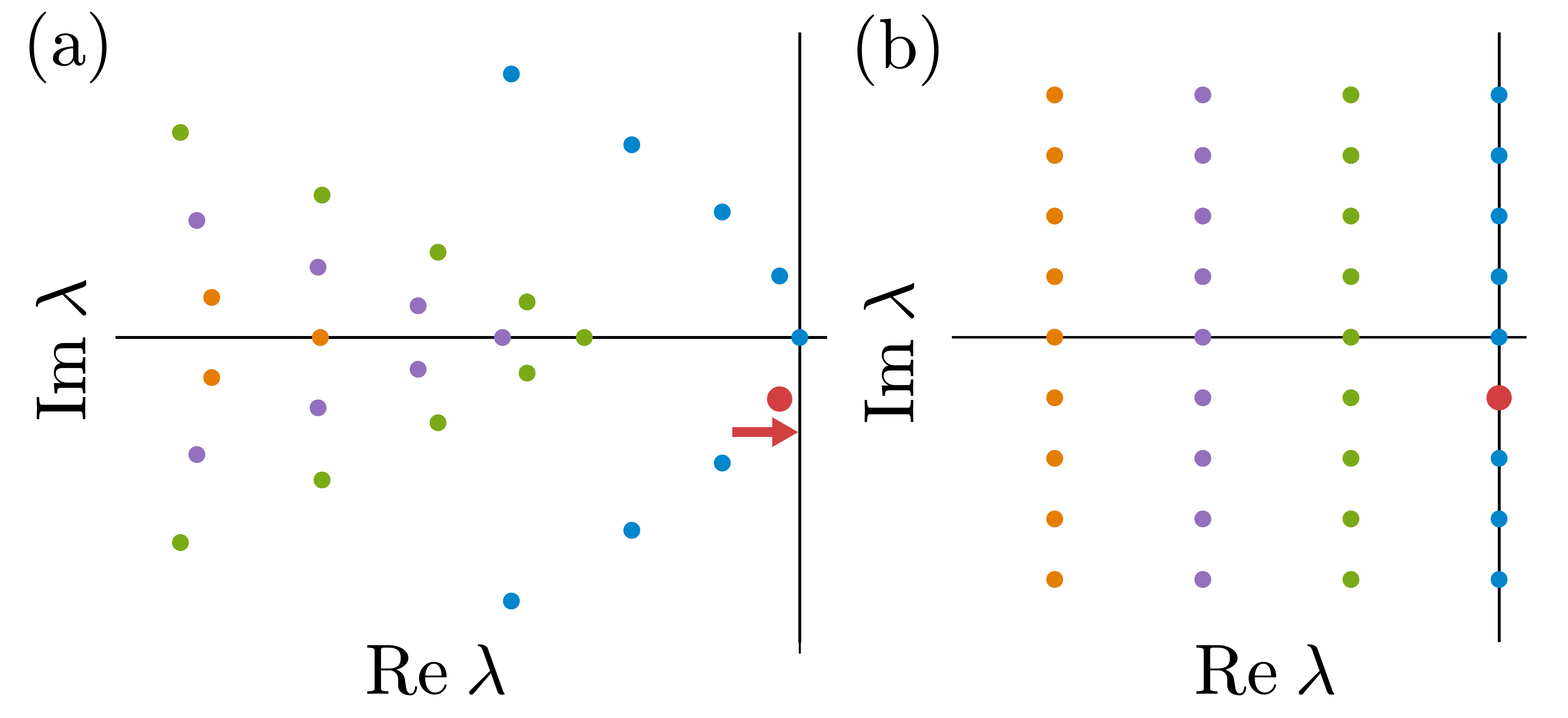}
		\caption{
			A schematic diagram for eigenvalues of $L_{\bm X}^*$, which converges to the classical limit. (a) Semiclassical regime. (b) Classical limit.
			}
		\label{fig_3}
	\end{center}
\end{figure}

Figure~\ref{fig_3} schematically shows the behavior of the eigenvalues of $L_{\bm X}^*$ approaching those of the deterministic system in the classical limit with a stable limit-cycle solution.
The eigenvalues in the classical limit are given in the form  $\lambda_c = m \kappa_1  + i n \omega_1~(m=0,1,2,\ldots$ and $n = 0, \pm 1, \pm2)$, where $\kappa_1$ is the real part of the largest non-zero eigenvalue 
and $\omega_1$ is the imaginary part of the pure-imaginary eigenvalue with the smallest absolute imaginary part.
In the example of a quantum van der Pol oscillator with quantum Kerr effect in Sec. IV, $ \kappa_1 = -\gamma_1$ and  $ \omega_1 = \Omega_c$.
As the system approaches the classical limit, each curved branch of eigenvalues $\Lambda_q$ of $L_{\bm X}$ in Fig.~\ref{fig_3}(a)
 approaches the corresponding straight branch of eigenvalues $\lambda_c$ in Fig.~\ref{fig_3}(b)
 in the classical limit.

The above formal correspondence with the conventional definition of the asymptotic phase in the classical limit supports the validity of our definition of the quantum asymptotic phase.

\section{Phase function of a quantum damped harmonic oscillator}

In Ref.~\cite{thomas2019phase}, Thomas and Lindner considered the stochastic phase function for a classical damped harmonic oscillator described by a multi-dimensional Ornstein-Uhlenbeck process.
In this section, generalizing their result, we consider a simple quantum harmonic oscillator with a damping and formally calculate the phase function.
This system is linear and does not possess a limit cycle in the classical limit, but the isochronous phase function as defined in Sec.~III can still be introduced (as the system lacks an asymptotic periodic orbit in this case, we do not use the term 'asymptotic'). 

The eigenoperator $V_1$ of the adjoint Liouville operator ${\mathcal L}^*$ can be analytically obtained in this case.
The evolution of a damped harmonic oscillator is described by a quantum master equation 
\begin{align}
	\label{eq:damp_ho}
	\dot{\rho}
	= {\mathcal L}\rho 
	= -i[ {\omega} a^{\dag} a, \rho] 
	+  \gamma \mathcal{D}[a]\rho,
\end{align}
where $\omega$ is the natural frequency of the system,
$\gamma$ denotes the decay rate for the linear damping, and $\mathcal{D}$ is the Lindblad form~\cite{carmichael2007statistical}.
The eigenoperator associated with the slowest non-vanishing decay rate of the adjoint Liouville operator ${\mathcal L}^*$ of ${\mathcal L}$ is simply given by $V_1 = a$, i.e., ${\mathcal L}^* a = \overline{ \Lambda _1}a$, 
where 
$\overline{\Lambda_1} = -\gamma/2 - i\omega$~\cite{barnett2000spectral, briegel1993quantum}.
Therefore, the phase function $\Phi_q(\bm{\alpha})$ of the coherent {state} $\bm{\alpha}$ is given by
\begin{align}
	\Phi_q(\bm{\alpha}) = \arg \langle \rho_t, a \rangle_{tr} = \arg \bra{\alpha} a \ket{\alpha} = \arg \alpha = \arg \left( re^{i\theta} \right) = \theta,
\end{align}
where $\alpha = r e^{i \theta}$, and the phase function $\Phi_q(\rho_t)$ of the density operator $\rho_t$ at time $t$ is given by
\begin{align}
	\Phi_q(\rho_t) = \arg \langle a \rangle_t = \arg \langle \rho_t, a \rangle_{tr}.
\end{align}
For the initial condition $\rho_0 = \ket{\alpha_0} \bra{\alpha_0}$
with $\alpha_0 = r_0 e^{i \theta_0}$, the expectation of $a$ evolves as
\begin{align}
\left. \frac{d}{dt} \langle a \rangle_t \right|_{t=t_0}
= \langle \dot\rho_t|_{t=t_0}, a \rangle_{tr} 
= \langle {\mathcal L} \rho_{t_0}, a \rangle_{tr} 
= \langle \rho_{t_0}, {\mathcal L}^* a \rangle_{tr} 
= \overline{\Lambda_1} \langle \rho_{t_0}, a \rangle_{tr} 
= \overline{\Lambda_1} \langle a \rangle_t,
\end{align}
which gives $\langle a \rangle_t = e^{\overline{\Lambda_1} t} \langle a \rangle_0 = e^{\overline{\Lambda_1} t} \bra{\alpha_0} a \ket{\alpha_0} = e^{\overline{\Lambda_1} t} \alpha_0
$. Thus, the phase of the state $\rho_t$ is given by
\begin{align}
\Phi_q(\rho_t) = \arg ( e^{ (-\gamma/2 - i \omega) t} \alpha_0) = -\omega t + \theta_0,
\end{align}
which decreases with a constant frequency $\omega$.

As shown in this example of a quantum damped harmonic oscillator, we can formally introduce the phase function for a wide class of oscillators, even if the system does not exhibit limit-cycle dynamics. In a special case where the system exhibits limit-cycle dynamics, our definition of the phase function plays the role of the quantum asymptotic phase. 


\end{document}